# Blind Decoding-Metric Estimation for Probabilistic Shaping via Expectation Maximization

Fabian Steiner[(1)], Patrick Schulte[(1)], Georg Böcherer[(2)]

[(1)] Technical University of Munich, {fabian.steiner, patrick.schulte}@tum.de
[(2)] Mathematical and Algorithmic Sciences Lab, Huawei France, georg.boecherer@huawei.com

**Abstract** *An unsupervised learning approach based on expectation maximization is proposed to obtain the parameters of a soft decision forward error correction decoding metric for probabilistic shaping. The algorithm depends only on the channel observations and does not require transmitted data.*

**Introduction**

Probabilistic amplitude shaping (PAS)[1] recently gained a lot of interest in the field of optical communications[2] to increase the spectral efficiency (SE) and flexibility of transceivers. Several metrics have been suggested to characterize the performance limits of a coded modulation system, e.g., generalized mutual information (GMI)[3], normalized generalized mutual information (NGMI)[4,5] and the cross entropy based uncertainty[6]. The essential component of these metrics is a simple and tractable model of the optical communication channel as seen by the FEC decoder.

A pragmatic and empirically accurate model is an additive white Gaussian noise (AWGN) channel. However, its parameters (e.g., noise variance, gain) must be estimated at the receiver. For probabilistic shaping (PS), the receiver also needs to know the transmit symbol distribution to achieve the best performance[7]. Usually, these parameters are either known or obtained by data-aided (DA) maximum likelihood (ML) estimation using both the transmitted and received data[2] (Sec. III-B). In general, the later approach needs to be used with caution, as it may lead to an over-estimation of achievable rates.

We propose an unsupervised learning approach based on expectation maximization (EM) that estimates all model parameters for a decoding metric suitable for PAS and soft-decision forward error correction (SD-FEC) schemes in a blind fashion. The suggested solution uses only the channel outputs and learns all relevant parameters on the fly. This is particularly important for PAS which is inherently rate flexible and a separate signaling of the modulation parameters is undesirable. We validate the approach by using recorded data from transmission experiments[2] and compare to DA estimation.

**Decoding Metric and Achievable Rates**

After coherent digital signal processing (DSP) the receiver obtains a length $n$ sequence $y^n$ of noisy QAM symbols, which is used to detect the transmitted sequence $x^n$. For this, the decoder uses a decoding metric $q^n(x^n, y^n) : \mathcal{X}^n \times \mathbb{C}^n \to \mathbb{R}^+$, where $\mathcal{X}$ is the employed constellation at the transmitter. The decoding decison is $\hat{x}^n$ if

$$\hat{x}^n = \underset{x^n \in \mathcal{C}}{\arg\max}\, q^n(x^n, y^n) \quad (1)$$

where $\mathcal{C}$ is the set of all forward error correction (FEC) codewords. Practical receivers use a memoryless metric, i.e.,

$$q^n(x^n, y^n) = \prod_{i=1}^n q(x_i, y_i). \quad (2)$$

The performance is characterized by a cross-entropy

$$\min_{s>0} u_\text{S} = \min_{s>0} \mathrm{E}\left[-\log_2\left(\frac{q(X,Y)^s}{\sum_{x\in\mathcal{X}} q(x,Y)^s}\right)\right]. \quad (3)$$

which is the uncertainty the FEC decoder needs to resolve[6]. Both the NGMI[5] and the achievable binary code (ABC) rate[6] are determined by the cross entropy (3). An achievable transmission rate is

$$R_\text{a} = [\mathrm{H}(X) - u_\text{S}]^+ \quad \text{[bits/channel use]}. \quad (4)$$

Depending on the FEC solution at hand (soft or hard input, symbol- or bit-wise decoding), the metric (3) can be instantiated appropriately. In the following, we concentrate on a bit-wise decoding metric for SD-FEC which is the relevant one for state-of-the-art codes, such as spatially coupled low-density parity-check (SC-LDPC) codes, decoded with belief propagation (BP). We associate each constellation point $x$ in the $M = 2^m$- ary constellation $\mathcal{X}$ with a binary label $\chi(x) = \boldsymbol{b} \in \{0,1\}^m$. Usually, a binary reflected Gray code (BRGC) is used. The FEC decoder inputs are the soft information values

$$l_j = \log\left(\frac{\sum_{x \in \mathcal{X}_j^0} p_{Y|X}(y|x)P_X(x)}{\sum_{x \in \mathcal{X}_j^1} p_{Y|X}(y|x)P_X(x)}\right) \quad (5)$$

for $j = 1, \ldots, m$ and $\mathcal{X}_j^b = \{x \in \mathcal{X} : [\chi(x)]_j = b\}$. For this choice, (3) can be written as

$$u_{\mathsf{S}} = \min_{s \geq 0} \sum_{j=1}^m \mathsf{E}\left[-\log_2\left(\frac{e^{(1-2B_j)(L_j s/2)}}{e^{-(L_j s)/2} + e^{(L_j s)/2}}\right)\right] \quad (6)$$

and the ABC rate[6] (aka NGMI[5]) is

$$R_{\mathsf{abc}} = 1 - \frac{1}{m} u_{\mathsf{S}} \quad (7)$$

i.e., there exists a binary rate $R_{\mathsf{abc}}$ code that can recover $x^n$ from $y^n$.

**Channel Model for the Decoding Metric**

We employ an AWGN model to derive a decoding metric $q(x,y)$ for a SD-FEC code

$$Y = \Delta \cdot X + N \quad (8)$$

where the channel input $X$ takes values in the discrete, $M = 2^m$-ary quadrature amplitude modulation (QAM) constellation $\mathcal{X}$ and is distributed according to a distribution $P_X$ with $P_X(x_j) = p_j, j = 1, \ldots, M$. In many cases, a Maxwell-Boltzmann (MB) distribution is a good proxy for the capacity achieving distribution, in which case $P_X$ is fully specified by a parameter $\nu$, i.e., $P_X^\nu(x_j) \propto e^{-\nu|x_j|^2}, j = 1, \ldots, M$. The noise term $N$ is assumed to be circularly symmetric complex Gaussian with zero mean and variance $\sigma^2$, i.e., $N \sim \mathcal{CN}(0, \sigma^2)$. The parameter $\Delta \in \mathbb{R}^+$ models the channel gain. The complete model is specified by the parameter vector $\boldsymbol{\theta} = (\Delta, \sigma^2, \boldsymbol{p})$ for a general $P_X$ or by $\boldsymbol{\theta} = (\Delta, \sigma^2, \nu)$ for an MB distribution $P_X$, respectively. The decoder models the optical channel by the AWGN channel probability density function (PDF)

$$p_{Y|X}(y|x;\boldsymbol{\theta}) = \frac{1}{\pi \theta_2} e^{-\frac{|y-\theta_1 x|^2}{\theta_2}}. \quad (9)$$

The model (8) can easily be extended to complex channel gains $\Delta \in \mathbb{C}$, or even constellation point dependent gains and noise variances to model scenarios where points with higher energy experience additional non-linear phase noise.

**PAS Parameter Estimation via Expectation Maximization**

Suppose we are given $n$ channel observations $y_1, y_2, \ldots y_n$. A typical ML problem formulation for the parameter vector $\boldsymbol{\theta}$ is

$$\begin{aligned}\boldsymbol{\theta}_{\mathsf{opt}} &= \operatorname*{argmax}_{\boldsymbol{\theta}} \sum_{i=1}^N \log\left(p_Y(y_i; \boldsymbol{\theta})\right) \\ &= \operatorname*{argmax}_{\boldsymbol{\theta}} \underbrace{\sum_{i=1}^N \log\left(\sum_{x \in \mathcal{X}} p_{Y|X}(y_i|x; \boldsymbol{\theta})P_X(x; \boldsymbol{\theta})\right)}_{L(\boldsymbol{\theta})}.\end{aligned} \quad (10)$$

Solving this problem directly by calculating the derivative with respect to $\boldsymbol{\theta}$ turns out to be challenging because of the latent variable $X$ in (8). The imposed channel model is a Gaussian mixture model (GMM), and problems of this kind have received renewed interest recently with regard to unsupervised learning algorithms. A standard approach is to use the iterative EM algorithm[8,9], which decomposes (10) into an expectation (E) and maximization (M) step. The $t$-th E-step computes an auxiliary distribution $Q_{X_i}^{(t)}$ as

$$Q_{X_i}^{(t)}(x) = \frac{p_{Y|X}(y_i|x; \boldsymbol{\theta}^{(t-1)})P_X(x; \boldsymbol{\theta}^{(t-1)})}{p_Y(y_i; \boldsymbol{\theta}^{(t-1)})}. \quad (11)$$

The M-step then maximizes the objective

$$\sum_{i=1}^n \sum_{x \in \mathcal{X}} Q_{X_i}^{(t)}(x) \log\left(p_{Y|X}(y_i|x; \boldsymbol{\theta})P_X(x; \boldsymbol{\theta})\right) \quad (12)$$

with respect to $\boldsymbol{\theta}$, which is a lower bound to $L(\boldsymbol{\theta})$[9]. The stationary points of the optimization in (12) are given as

$$\Delta_{\mathsf{opt}}^{\mathsf{EM}} = \frac{\sum_{i=1}^n \sum_{x \in \mathcal{X}} Q_{X_i}^{(t)}(x) \Re(y_i x^*)}{\sum_{i=1}^N \sum_{x \in \mathcal{X}} Q_{X_i}^{(t)}(x) |x|^2}, \quad (13)$$

$$\sigma_{\mathsf{opt}}^{2,\mathsf{EM}} = \frac{1}{n} \sum_{i=1}^n \sum_{x \in \mathcal{X}} Q_{X_i}^{(t)}(x) \left|y_i - \Delta_{\mathsf{opt}}^{\mathsf{EM}} x\right|^2, \quad (14)$$

$$p_{j,\mathsf{opt}}^{\mathsf{EM}} = \frac{1}{n} \sum_{i=1}^n Q_{X_i}^{(t)}(x_j) \quad (15)$$

where $\boldsymbol{p}_{\mathsf{opt}}^{\mathsf{EM}} = (p_{1,\mathsf{opt}}^{\mathsf{EM}}, \ldots, p_{M,\mathsf{opt}}^{\mathsf{EM}})$. If we assume an MB distribution on $\mathcal{X}$, the value for $\nu_{\mathsf{opt}}$ is the solution of the non-linear equation:

$$\frac{1}{n} \sum_{i=1}^n \mathsf{E}_{X \sim Q_{X_i}^{(t)}}\left[X^2\right] = \mathsf{E}_{X \sim P_X^{\nu_{\mathsf{opt}}}}\left[X^2\right]. \quad (16)$$

The EM algorithm is initialized with a reason-able choice for the initial value $\theta^{(0)}$ and the M- and E-steps are run for a given number of iterations.

**Numerical Results**

As shown in [9], EM does not necessarily converge to the globally optimal solution of (10), but usually only to a local optimum. To guarantee convergence to good parameter values, the EM algorithm needs to be initialized carefully. For this, we choose the initial parameter $\theta^{(0)}$ by a modified version of K-Means[8], which only uses the channel outputs and includes a constraint on the equispaced constellation points. The number of initial clusters for K-Means is an important parameter, as not all of the $M$ constellation points might have been transmitted. We circumvent this problem by initializing K-Means with a smaller number of clusters. For instance, for 64-QAM, we run the EM algorithm with initializations obtained from K-Means with the number of clusters set to $\{4, 16, 36, 64\}$ and choose the parameter set that minimizes the uncertainty (6). This corresponds to the common practice of initializing EM with different random starting points and using the set of parameters maximizing the objective (10). We use a DA scheme[2] as a reference and compare it to our proposed EM approach. For the DA scheme, the receiver has full access to the sequence pairs $(x^n, y^n)$ and the estimated parameters are

$$\Delta_{\text{opt}}^{\text{DA}} = \sum_{i=1}^{n} \frac{\Re(y_i x_i^*)}{|x_i|^2}, \quad \sigma_{\text{opt}}^{2,\text{DA}} = \frac{1}{n}\sum_{i=1}^{n}\left|y_i - \Delta_{\text{opt}}^{\text{DA}} x_i\right|^2,$$

$$p_{\text{opt},j}^{\text{DA}} = \frac{|\{i \in \{1,\ldots,n\}|x_i = x_j\}|}{n}, j=1,\ldots,M.$$

To assess the performance of both schemes, we use the recorded sequences of one of our previous transmission experiments[2], in which four shaping modes with different entropies were investigated. The length of each sequence was $n \approx 20\,000$ QAM symbols. Mode 4 effectively corresponds to a 36-QAM constellation. The results are shown in Fig. 1. We observe that both the DA and EM approaches achieve the same achievable rates, showing that EM can accurately estimate the parameters for all considered modes. EM converged in less than 30 iterations.

**Conclusions**

We proposed an unsupervised learning approach based on EM to estimate the decoding metric for probabilistic shaping. The suggested solution operates blindly and requires only access to the observations of the channel outputs. Numerical results based on experimental data with 64-QAM and different shaped modes show that the performance matches that of a DA scheme.

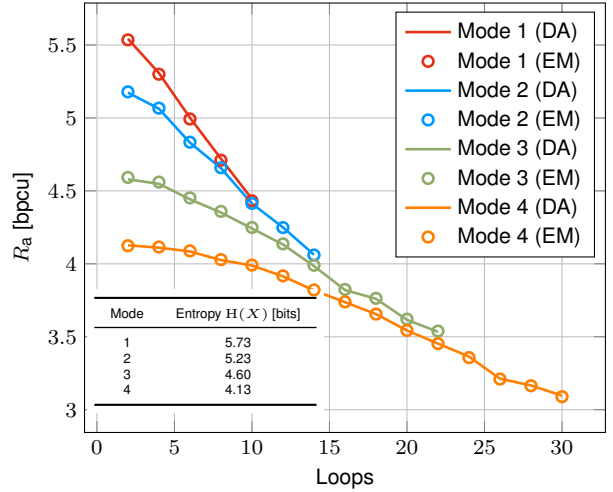

**Fig. 1:** Achievable rates obtained via data-aided (DA) and blind (EM) parameter estimation. The sequences are taken from the transmission experiment in[2]. One loop corresponds to 240 km.